\documentclass[10pt,twocolumn,numbers]{article}

\usepackage{graphicx}

\usepackage{amssymb,amsthm,amsmath,dsfont,bbm,mathtools,stmaryrd}

\usepackage{lineno}

\usepackage[colorlinks=true, citecolor=blue]{hyperref}
\usepackage[lined,ruled,commentsnumbered]{algorithm2e}
\usepackage{algcompatible}

\usepackage{wrapfig}
\usepackage[numbers,sort&compress]{natbib}
\usepackage{comment}
\usepackage{longtable}
\usepackage{multirow}
\usepackage{bbding}
  
\usepackage[bottom]{footmisc}
\usepackage{booktabs}
\usepackage{pifont}
\usepackage{float}
\usepackage{wrapfig}
\usepackage{framed,xcolor}
\usepackage{newfloat}
\usepackage[xcolor,framemethod=TikZ]{mdframed}
\usepackage{amsthm}
\usepackage{thmtools}
\usepackage{authblk}
\patchcmd{\thebibliography}{\section*{\refname}}{}{}{}
\usepackage{tikz}
\usetikzlibrary{arrows,shapes,chains,calc,patterns,decorations.pathreplacing}
\usepackage{stmaryrd}
\usepackage{fancyhdr}
\usepackage{subcaption}
\usepackage{comment}
\def\BibTeX{{\rm B\kern-.05em{\sc i\kern-.025em b}\kern-.08em
		T\kern-.1667em\lower.7ex\hbox{E}\kern-.125emX}}

\renewcommand{\headrulewidth}{2pt}

\newlength\FHoffset

\fancyheadoffset{\FHoffset}

\newlength\FHleft
\newlength\FHright

\newbox\FHline
\setbox\FHline=\hbox{\hsize=\paperwidth%
	\hspace*{\FHleft}%
	\rule{\dimexpr\headwidth-\FHleft-\FHright\relax}{\headrulewidth}\hspace*{\FHright}%
}

\newtheoremstyle{theoremdd}
{\topsep}
{\topsep}
{\itshape}
{0pt}
{\fontfamily{cmss}\selectfont\bfseries}
{.}
{ }
{\thmname{#1}\thmnumber{ #2}\thmnote{ (#3)}}

\theoremstyle{theoremdd}

\mdfdefinestyle{myStyle}{roundcorner=5pt,backgroundcolor=brown!20,linecolor=red!40!black,linewidth=1.5pt}

\usepackage{titlesec}
\titleformat*{\section}{\fontfamily{cmss}\selectfont\large\bfseries\color{red!40!black}}
\titleformat*{\subsection}{\fontfamily{cmss}\selectfont\normalsize\bfseries\color{red!40!black}}
\titleformat*{\subsubsection}{\fontfamily{cmss}\selectfont\normalsize\color{red!40!black}}
\usepackage[left=0.69 in, right=0.69 in, top=1.1 in, bottom=1.3 in]{geometry}
\graphicspath{{./Figures/}}

\renewcommand\abstractname{\fontfamily{cmss}\selectfont\normalsize\bfseries\color{red!40!black}\textbf{Abstract}}

\renewenvironment{abstract}{%
	\centering\small
	\list{}{\leftmargin1.5cm \rightmargin\leftmargin}
	\item\relax
	
	\begin{mdframed}[]
		\item[\hskip\labelsep\scshape\abstractname.]%
	}{%
	\end{mdframed}
	\endlist \par\bigskip
}
\makeatletter
\patchcmd{\@maketitle}{\LARGE \@title}{\fontfamily{cmss}\selectfont\LARGE\color{red!40!black}\@title}{}{}
\makeatother

\begin{document}

		
		
		\title{Generalized adaptive smoothing based neural network architecture for traffic state estimation}
		
			

\author[1]{Chuhan Yang$^{\star}$}
\author[2]{Ambadipudi Sai Venkata Ramana}
\author[1,2]{Saif Eddin Jabari}

\affil[1]{New York University Tandon School of Engineering, Brooklyn NY, U.S.A.}
\affil[2]{New York University Abu Dhabi, Saadiyat Island, P.O. Box 129188, Abu Dhabi, U.A.E.}

\date{}


\twocolumn[
\begin{@twocolumnfalse}
	
\maketitle	

\begin{abstract}
	The adaptive smoothing method (ASM) is a standard data-driven technique used in traffic state estimation. The ASM has free parameters which, in practice, are chosen to be some generally acceptable values based on intuition. However, we note that the heuristically chosen values often result in un-physical predictions by the ASM.
In this work, we propose a neural network based on the ASM which tunes those parameters automatically by learning from sparse data from road sensors. We refer to it as the adaptive smoothing neural network (ASNN). We also propose a modified ASNN (MASNN), which makes it a strong learner by using ensemble averaging. The ASNN and MASNN are trained and tested two real-world datasets. Our experiments reveal that the ASNN and the MASNN outperform the conventional ASM.
	
\end{abstract}
\bigskip
\end{@twocolumnfalse}
]

		
	
	
	

\section{INTRODUCTION}

Macroscopic traffic state variables such as flow rate and average vehicle speed, as key measurements of traffic conditions on road segments in a traffic network, have been instrumental in transportation planning and management. The accurate estimation of traffic state variables has received considerable attention because traffic state variables cannot be directly measured everywhere due to technological and financial limitations, and need to be estimated from noisy and incomplete traffic data. 
As a result, the process of the inference of traffic state variables from partially observed traffic data, which is referred to as \emph{Traffic state estimation} (TSE), has been the subject of much systematic investigation. Commonly, traffic data comes from many heterogeneous sources including stationary detector data (SDD) collected by sensors fixed in the infrastructure or floating-car data (FCD) collected by GPS devices and cellphones.

The approaches of interest are often grouped into model-driven and data-driven \cite{seo2017traffic}. Model-driven approaches like Lighthill-Whitham-Richards (LWR) model \cite{lighthill1955kinematic}, \cite{richards1956shock},
 Aw-Rascle-Zhang (ARZ) model \cite{aw2000resurrection} use hyperbolic conservation laws coupled with dynamical measurement models to estimate traffic states. Data-driven approaches use supervised learning algorithms to predict the traffic states.  A widely used data-driven approach to estimate the traffic state is the adaptive smoothing method (ASM) \cite{treiber2002reconstructing}.  Other data-driven approaches used in TSE include deep neural networks \cite{jia2016traffic}, \cite{benkraouda2020traffic}, \cite{thodi2022incorporating}, support vector regression \cite{xiao2018speed} and matrix factorization \cite{li2022nonlinear}.

The ASM is widely used in traffic state estimation because of its simplicity in implementation and low computational cost.
It is basically an interpolation method that reconstructs the spatio-temporal traffic state by passing a nonlinear low-pass filter to discrete traffic data to produce missing traffic state variables as smooth functions of space and time. Specifically, the ASM accounts for the fact that perturbations in traffic flow propagate downstream in free-flow and upstream in congestion with a characteristic constant velocity, and estimates the traffic states as superpositions of two linear anisotropic lowpass filters with an adaptive weight factor.

A major disadvantage of ASM is the free parameters in the method. Ideally, one would determine the free-parameters by calibrating them to match real data. However, in most cases the real data available is so sparse and scattered that conventional fitting methods are not efficient. This leads to non-physical traffic state predictions by the method.

To address this shortcoming, we develop a neural network architecture based on ASM which we call ASNN. 
We believe the proposed ASNN allows for better parameter estimation from sparse data. We first develop a vanilla version of the ASNN and then build a more sophisticated version, MASNN, by adding multiple smoothing nodes with different initializations and incorporating multiple a priori estimates. The MASNN  allows for finding suitable parameters for ASM where no easy access exists to prior information for the parameters. We compare ASM and our proposed method's performance on data from the Next Generation Simulation Program (NGSIM) and the Germany Highway Drone (HighD) dataset with different input settings.


The rest of the paper is organized as follows: Section \ref{sec:related_work} reviews the relevant literature. In Section \ref{sec:method}, we present the ASM and our proposed method ASNN, we also discuss ASNN's workflow and the generalization of ASNN, MASNN. The experiment details are illustrated in Section \ref{sec:exp_detail}, followed by a brief discussion of results in Section \ref{sec:discuss}. Finally, we draw our conclusion and discuss future research in Section \ref{sec:conclusion}.


\section{RELATED WORK}\label{sec:related_work}

The adaptive smoothing method (ASM) belongs to the group of data-driven methods known as structured-learning methods, which honor physics constraints such as conservation laws in order to improve the performance or interpretability, see \cite{thodi2022incorporating}, \cite{thodi2021learning}, \cite{shi2021physics}, \cite{huang2020physics}, \cite{schreiter2010two}, \cite{chen2018adaptive}. The physical constraints are incorporated in the model fitting stage, e.g., \cite{shi2021physics}, \cite{huang2020physics}
or infused in the model architecture, e.g., \cite{jabari2018stochastic}, \cite{jabari2018learning}, \cite{jabari2020sparse}, \cite{thodi2022incorporating}. Structured-learning methods have shown robust estimation performance and require limited data in the function fitting process. ASM is a structured-learning method because the interpolation process in ASM considers free-flow and congested traffic wave propagation characteristics. 

Different attempts exist in the literature regarding improvement and modification of the conventional ASM. For instance, \cite{schreiter2010two} discretized the ASM and applied Fast Fourier Transform techniques to further reduce the computation time for real-time applications. In \cite{chen2018adaptive}'s work,
the smoothing width parameters of the ASM dynamically change in a rolling-
horizon framework to capture complicated traffic states. But this work ignored other ASM parameters (free-flow and shock wave
speeds, transition width, critical speed) and their values are fixed.
The method proposed in our paper overcomes this limitation
by systematically searching for optimal parameters considered in ASM. Anisotropic kernels of the ASM are also applied in designing efficient
convolutional neural networks for traffic speed estimation in \cite{thodi2022incorporating}. A recent work of \cite{yang2022generalized} investigated ASM from a matrix completion perspective and proposed a systematic procedure to calculate
the optimal weight factors instead of using the conventional heuristic weight factors.


\section{METHODOLOGY}\label{sec:method}

\subsection{Notation}

We denote matrices using uppercase bold Roman letters ($\mathbf{W},\mathbf{Z},\mathbf{V}$,etc). We specify $\mathbf{J}$ as the all-ones matrix, in which every element is equal to one. The symbol $\odot$ represents the Hadamard product. $\|\cdot\|_{\mathrm{F}}$ is the Frobenius norm of a matrix. The symbol $*$ represents convolution operation. Other notations are described in Table \ref{table_parameter}.

\subsection{Conventional Adaptive Smoothing Method (ASM)}\label{sec:ASM}
The adaptive smoothing method takes any macroscopic traffic quantity $\mathbf{Z}$ as input. The elements of $\mathbf{Z}$, $\mathbf{Z}(x,t)$, can can represent traffic flux at space-time indices $(x,t)$ denoted $q(x,t)$, speeds ($v(x,t)$), or traffic densities ($\mathbf{\rho}(x,t)$). ASM calculates the estimated field based on the assumption that traffic is in one of two regimes: In free-flow traffic, perturbations essentially move downstream of the flow whereas the perturbations move upstream in the congested traffic. According to kinematic wave theory, the perturbations travel with a constant wave speed along the characteristic curves over space and time in each of these regimes. ASM first calculates two a priori fields as follows:

\begin{equation} 
    \label{eqn:fields}
    \begin{aligned}
     &\mathbf{Z}^{\rm free} (x,t) = \frac{1}{N(x,t)} \sum_n \phi \left( x-x_n, t-t_n-\frac{x-x_n}{c_{\rm free}} \right) z_n, \\
     &\mathbf{Z}^{\rm cong} (x,t) = \frac{1}{N(x,t)} \sum_n \phi \left( x-x_n, t-t_n-\frac{x-x_n}{c_{\rm cong}} \right) z_n,
    \end{aligned}
\end{equation}

where $c_{\rm free}$ is the characteristic wave speed in free-flow regime and $c_{\rm cong}$ is the characteristic wave speed in the congested regime and $N(x,t)$ is a normalization constant, $\phi(\cdot,\cdot)$ is the smoothing kernel function:
\begin{equation}\label{eqn:phi}
    \phi(x,t) = \exp\Big(-\frac{|x|}{\sigma}-\frac{|t|}{\tau}\Big).
\end{equation}
$(x_n,t_n)$ are discrete position and time pairs
at which the data ($z_n$) are measured. $\sigma$ and $\tau$ represent the range of spatial smoothing in $x$ and temporal smoothing in $t$.
We can also apply other localized functions such as a two-dimensional Gaussian kernels. 
The output will be a superposition of these two a priori estimates with an adaptive weight field $\mathbf{W}^{\rm cong}$ is written in a compact form as
\begin{equation}
\label{eqn:z_convex}
    \begin{aligned}
     \mathbf{Z} = \mathbf{W}^{\rm cong} \odot \mathbf{Z}^{\rm cong} + (\mathbf{J}-\mathbf{W}^{\rm cong}) \odot \mathbf{Z}^{\rm free}.
    \end{aligned}
\end{equation}

$\mathbf{W}^{\rm cong} \in [0,1]$ also depends nonlinearly on the two a priori estimates, given by:
\begin{equation}
\label{eqn:pre-weight}
\begin{aligned}
\mathbf{W}^{\rm cong} = \frac{1}{2} \left[ 1+\tanh\left( \frac{Z_{\rm thr}-\min \{ \mathbf{Z}^{\rm free}, \mathbf{Z}^{\rm cong} \} }{\Delta Z} \right) \right].
\end{aligned}
\end{equation}
Note the operators $\min\{\cdot,\cdot\}$ and $\tanh(\cdot)$ in Eq.\eqref{eqn:pre-weight} are applied elementwise.
 $Z_{\rm thr}$ is the critical threshold, $\Delta Z$ is the transition range between congested and free traffic. 
 

\begin{figure*}
\begin{center}
\includegraphics[width=0.95\textwidth]{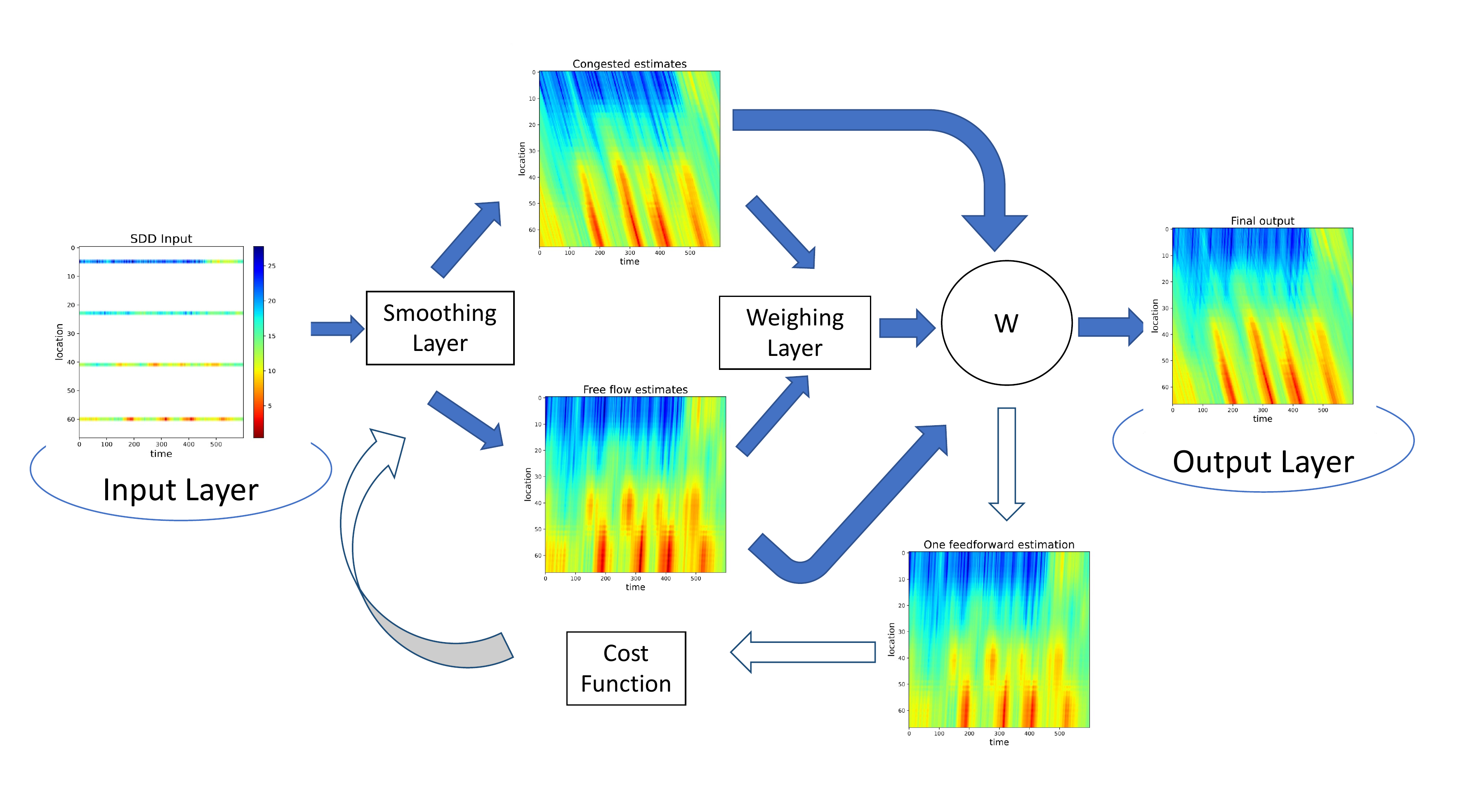}    
\caption{Adaptive Smoothing Neural Network for Traffic State estimation} 
\label{fig:ASNN_flow}
\end{center}
\end{figure*}
 
 There are six parameters involved in ASM estimation. The typical values of ASM parameters and their interpretation are summarized in Table \ref{table_parameter}. \cite{treiber2011reconstructing} suggested 
 that suitable values for $\sigma$ and $\tau$ are half of the inter-detector spacing and half of the sampling time respectively. Further, they performed a sensitivity analysis by visual inspection and noted that the ASM is less sensitive to remaining parameters. However, we noted in our experiments and also it can be observed from figure 3 in
 \cite{treiber2011reconstructing} that the ASM is sensitive to parameter values and we expect that the sensitivity is due to $Z_{thr}$ and $\Delta Z$. This demands for a robust algorithm to optimize the ASM parameters. Moreover, there are cases where we have no prior knowledge of the parameter values to start with and a method of tuning ASM parameters is essential. For example, we don't know the smoothing width value with mixture signals input recorded by heterogeneous sources. 
To this end, we formulate ASM as a neural network (ASNN) to efficiently handle the tuning of the field parameters. Our architecture can also accommodate multiple wave speeds, which allows one to determine optimal parameters setting when
typical values as those given in Table \ref{table_parameter} do not
work.


\begin{table}[h]
\small
\caption{Parameter Settings}
\label{table_parameter}
\begin{center}
\begin{tabular}{|c|c|l|}
\hline
Parameter & Value & Description\\
\hline
$c_{\rm cong}$ & -15 km/h & Congested wave speed\\
$c_{\rm free}$ & 80 km/h & Free-flow wave speed\\
$v_{\rm thr}$ & 60 km/h & Critical traffic speed\\
$\Delta v$ & 20 km/h & Transition width\\
$\sigma$ & $\Delta x/2$ & Space coordinate smoothing width\\
$\tau$ & $\Delta t/2$ & Time coordinate smoothing width\\
\hline
\end{tabular}
\end{center}
\end{table}

\subsection{Adaptive smoothing neural network (ASNN)}
Our adaptive smoothing neural network (ASNN) 
built using the ASM. It is, thus, designed to respect the kinematic wave theory inherently in the hidden layers as explained below.

ASNN is illustrated graphically in Fig.\ref{fig:ASNN_flow}.
The ASNN consists of the following layers: (i) An input layer where  stationary detector or heterogeneous source data are fed into the neural network. (ii) An output layer where a estimated field is reconstructed and given as output. (iii) Hidden layers, where information is processed between the input and output layers while respecting kinematic wave theory. From left to right in the hidden layers in Fig.\ref{fig:ASNN_flow}, the first hidden layer is the smoothing layer. In the smoothing layer, the input is fed into smooth spatio-temporal functions Eq.\eqref{eqn:phi} to generate an estimate of $\mathbf{Z}^{\rm free}$ and $\mathbf{Z^{\rm cong}}$ as given in Eq.\eqref{eqn:fields}.
The second hidden layer is the 
weighting layer. $\mathbf{Z}^{\rm free}$ and $\mathbf{Z^{\rm cong}}$ from the smoothing layer are used to calculate the weight matrix $\mathbf{W}^{\rm cong}$ according to \eqref{eqn:pre-weight}. $\mathbf{W}^{\rm cong}$ is then used in the convex combination \eqref{eqn:z_convex} of the $\mathbf{Z}^{\rm free}$ and $\mathbf{Z^{\rm cong}}$ to predict the output in the output layer. This architecture strictly preserves the workflow of ASM, thus when we use the same parameter values from Table \ref{table_parameter} as initialization, one feedforword of ASNN would reproduce the results of the conventional ASM. 


The cost function measures the average estimation accuracy for the neural network. In each training iteration, weights in neurons are adjusted to minimize the cost. In this work, we propose two cost functions
for ASNN: 

1.  Convolution cost function: We apply the $\ell_1$ regularized root mean square error based on the heuristic that the state of every cell in the estimated matrix should be close to those of its neighbors. The cost function is

\begin{equation}
\label{eqn:cost_fun}
\begin{aligned}
 \frac{\|\mathsf{P}_\Omega(\mathbf{Z}^{\rm est} - \mathbf{Z}^{\rm true})\|_F}{\|\mathsf{P}_\Omega(\mathbf{J}) \|_F} + \lambda \frac{\sum_{(x,t)}|\omega * \mathbf{Z}^{\rm est}(x,t)|}{{\|\mathbf{J} \|_F}}
\end{aligned}
\end{equation}

The binary mask operator $\mathsf{P}_\Omega$ evaluates the objective function only at the observed indices $\Omega$. $\lambda$ is the regularization parameter (or penalty parameter). The penalty term is a kernel convolution operation on the estimated matrix $\mathbf{Z}^{\rm est}$. $\omega$ is a filter kernel that takes two inputs that represent distance, $-a\leq dx\leq a$ and $-b\leq dy \leq b$, and produces a weight. For $|dx|>a$ or $|dy|>b$, $\omega(dx,dy)=0$. The expression of the convolution is:
\begin{multline}
\omega * \mathbf{Z}^{\rm est}(x,t) \\= \sum_{dx=-a}^a\sum_{dy=-b}^b \omega(dx,dy)\mathbf{Z}^{\rm est}(x-dx,y-dy)
\end{multline}

It measures the magnitude of the difference between each cell's estimated value and its neighbors' estimated values. 

\begin{figure}
     \centering
     \begin{subfigure}[b]{0.35\textwidth}
         \centering
         \includegraphics[width=\textwidth]{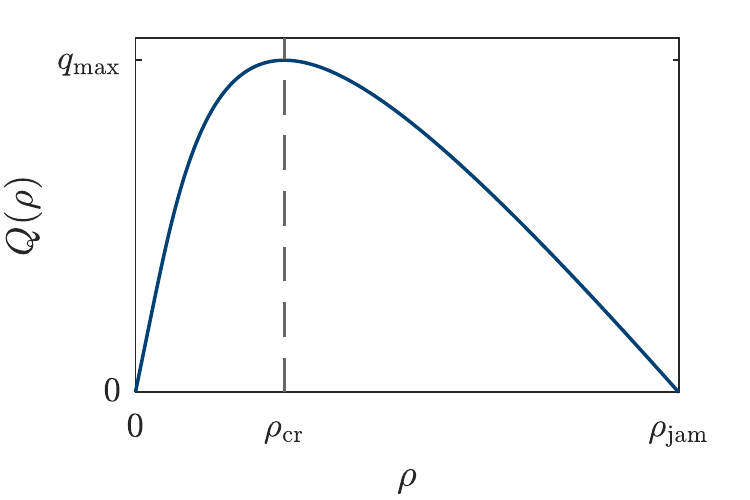}
         \caption{$Q(\rho)$}
         \label{fig:Q}
     \end{subfigure}
     \hfill
     \begin{subfigure}[b]{0.35\textwidth}
         \centering
         \includegraphics[width=\textwidth]{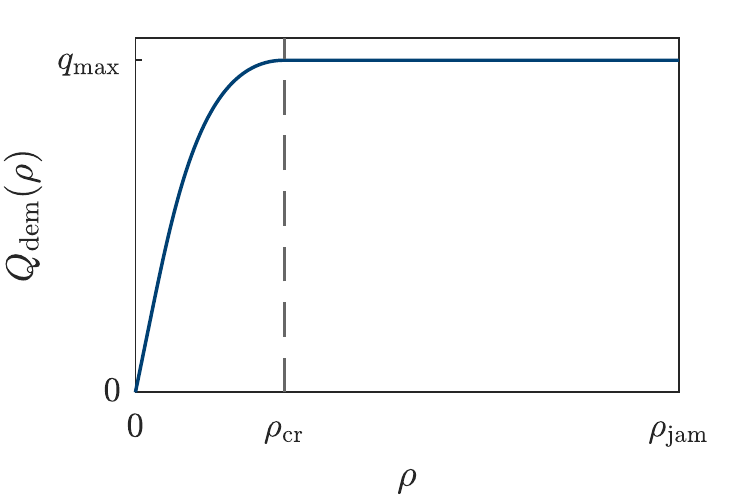}
         \caption{$Q_{\rm dem}(\rho)$}
         \label{fig:Qdem}
     \end{subfigure}
     \hfill
     \begin{subfigure}[b]{0.35\textwidth}
         \centering
         \includegraphics[width=\textwidth]{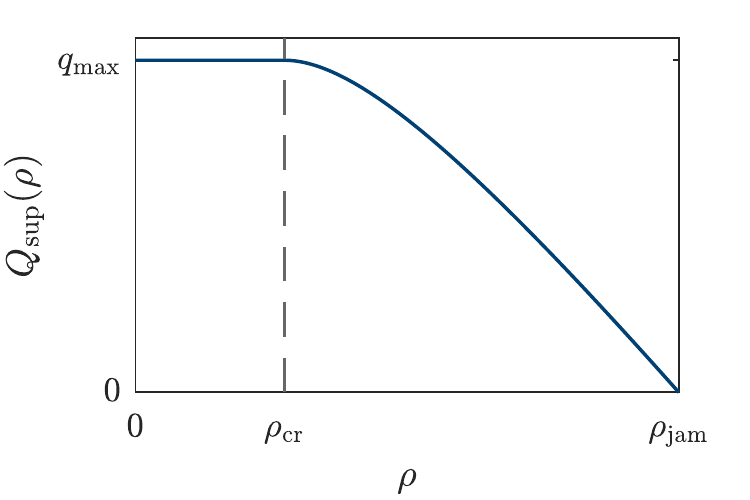}
         \caption{$Q_{\rm sup}(\rho)$}
         \label{fig:Qsup}
     \end{subfigure}
        \caption{Flux, demand, and supply relations. $\rho_{\rm jam}$ is the `jam density', which is a maximum traffic density at which vehicles cease to move, i.e., their speeds are zero, which results in zero flux.}
        \label{fig:Qs}
\end{figure}

2. Physics-informed cost function: We can encode traffic flow models as a regularization term in the cost function, which encourages the estimated result to follow some physics-based constraint. In this work, we consider a simple conservation principle based on the Lighthill-Witham-Richards (LWR) traffic flow model \cite{lighthill1955kinematic}, \cite{richards1956shock}: 
\begin{equation}
    \partial_t \rho(x,t) = -\partial_x Q\big(\rho(x,t) \big),
\end{equation}
where $Q(\cdot)$ is a concave flux function that describes the local rate of flow ($q(x,t)$) as a function of traffic density ($\rho(x,t)$). See in Fig.~\ref{fig:Q} for illustration.

LWR-type models are non-linear hyperbolic partial differential equations (PDEs), written in a conservative form \cite{leveque1992numerical}. A Godunov scheme developed to numerically solve the hyperbolic PDE results in the following discrete dynamical system:
\begin{equation}\label{eqn:LWR-PDE}
     \rho(x+\Delta x,t)  =  \rho(x,t)+\frac{\Delta x}{\Delta t}[q_{j-1\to j} - q_{j\to j+1}],
\end{equation}
where $q_{j-1\to j}$ and $q_{j\to j+1}$ denote the numerical traffic flux (the number of vehicles moving from cell $j-1$ to cell $j$ and from cell $j$ to cell $j+1$, respectively). Numerical traffic flux is calculated using the minimum supply-demand principle \cite{van2019traffic}:

\begin{equation}
    q_{j-1\to j} = \min{\big\{Q_{\text{dem}}(x,t-\Delta t), Q_{\text{sup}}(x,t)\big\}}
\end{equation}

\begin{equation}
    q_{j\to j+1} = \min{\big\{Q_{\text{dem}}(x,t),Q_{\text{sup}}(x,t+\Delta t)\big\}}
\end{equation}
The demand and supply functions in cell $(x,t)$ are defined as
\begin{equation}
  Q_{\text{dem}}(x,t) =
  \begin{cases}
    Q(\rho(x,t))       & \quad \text{if } \rho(x,t)\leq\rho_{\rm cr}\\
    q_{\text{max}}  & \quad \text{otherwise } 
  \end{cases}   
\end{equation}
and
\begin{equation}
    Q_{\text{sup}}(x,t) =
  \begin{cases}
    Q(\rho(x,t))       & \quad \text{if } \rho(x,t)>\rho_{\rm cr}\\
    q_{\text{max}}  & \quad \text{otherwise } 
  \end{cases},
\end{equation}
respectively. $\rho_{\rm cr}$ is the critical traffic density and $q_{\max} = Q(\rho_{\rm cr})$ is the maximal local flux. These local demand and supply relations are illustrated in Fig.~\ref{fig:Qdem} and Fig.~\ref{fig:Qsup}, respectively.

We use Eq.\eqref{eqn:LWR-PDE} as a regularization term 
\begin{equation}
 l_{\text{P}} = \sum_{(x,t)}\|\rho(x+\Delta x,t)  -  \rho(x,t)-\frac{\Delta x}{\Delta t}[q_{j-1\to j} - q_{j\to j+1}]\|_2 
\end{equation} 
and obtain the physics-informed cost function as follows:
\begin{equation}\label{eqn:phy-cost}
    \frac{\|\mathsf{P}_\Omega(\mathbf{Z}^{\rm est} - \mathbf{Z}^{\rm true})\|_F}{\|\mathsf{P}_\Omega(\mathbf{J}) \|_F} + \lambda \frac{l_{\text{P}}}{{\|\mathbf{J} \|_F}}.
\end{equation}
It promotes solutions that satisfy the discrete conservation equation \eqref{eqn:LWR-PDE}.

\subsection{ASNN workflow and MASNN}
To evaluate the model, we split the ground truth traffic state dataset into a training set and a testing set, where the training set consists of the partially observed traffic data (SDD or FCD). The training set is fed into the neural network as input. The cost function averages loss over the entire training set, and computes its gradients with respect to the parameters using backpropagation. The gradients are used to update each parameter. The training process takes multiple epochs to converge. In the event of no change in total cost, a maximum iteration number concludes the learning process.

\begin{figure*}
\includegraphics[width=\textwidth]{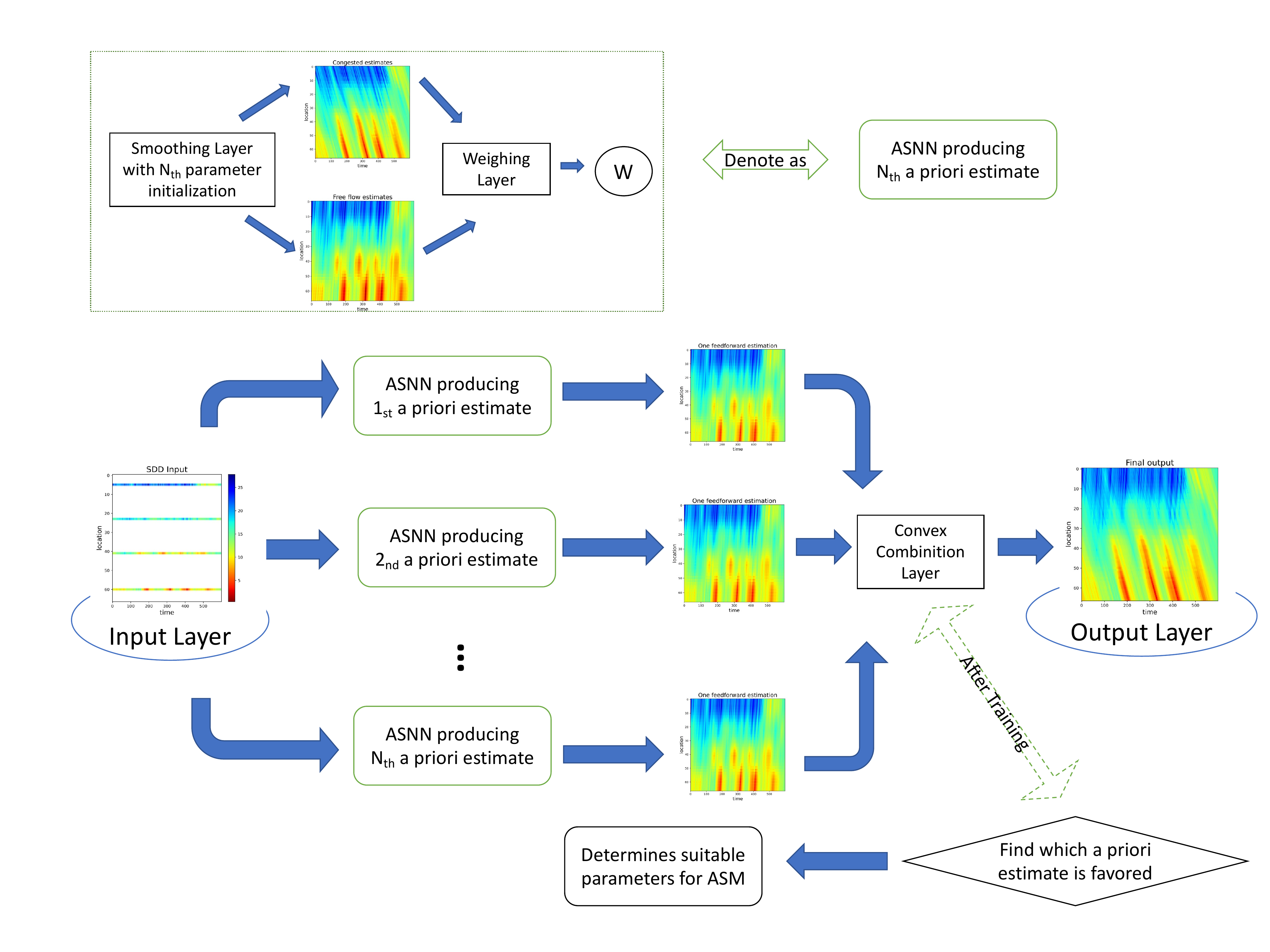}    
\caption{Multiple A Priori Adaptive Smoothing Neural Network for Traffic State estimation} 
\label{fig:MASNN-architecture}
\end{figure*}

ASNN inherits its architecture from ASM; it consists of the same six parameters as weights. We should point out that similar to other gradient-based learning methods, the ASNN architecture also faces the challenge of the diminishing gradient effects and weights in smoothing layers ($c,\sigma,\tau$) receive only minor changes in each update during training. This phenomenon indicates that parameters from smoothing layers require reasonable initialization for overall efficiency. Fortunately, these parameters are shared by ASNN and ASN and have the same interpretation, we can use their typical values as initialization and freeze the smoothing layers for fast convergence.

We noted from our experiments that the ASNN is a weak learner. For instance, when the data is hybrid
 i.e., when it comes from heterogeneous sources including stationary detectors and floating cars, the ASNN estimations are sensitive to the initialization. To alleviate this problem and to make the ASNN a robust learner, we develop the multiple a priori adaptive smoothing neural network (MASNN).
 MASNN is graphically demonstrated in Fig. \ref{fig:MASNN-architecture}.

 The MASNN basically uses the ensemble averaging technique to make the predictions independent of initial guess. To that effect, we consider an ensemble of ASNNs with different possible initializations and a convex combination layer is added to combine all ASNN estimations produced by different initializations. The weights of the ASNN and the convex combination layer are determined 
 together by optimizing the cost function.


\section{EXPERIMENTAL SETUP}\label{sec:exp_detail}
\begin{figure}
\begin{center}
\includegraphics[width=.5\textwidth]{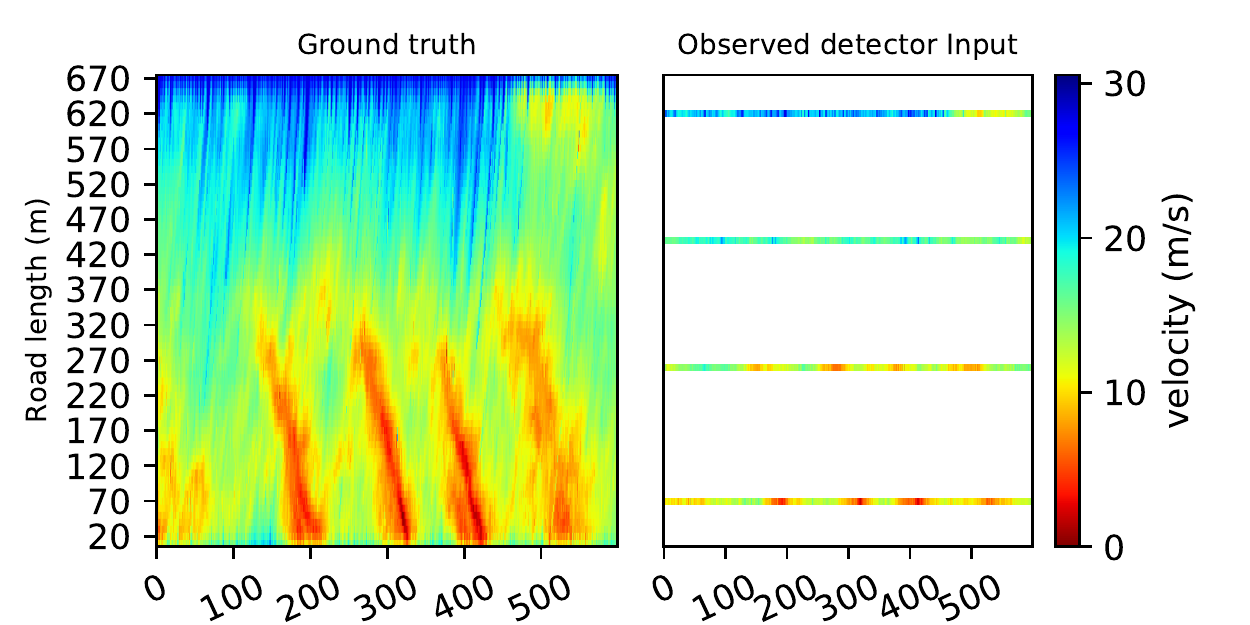}    
\caption{US101 data with stationary detector inputs} 
\label{fig:usdata}
\end{center}
\end{figure}

\begin{figure}
\begin{center}
\includegraphics[width=.5\textwidth]{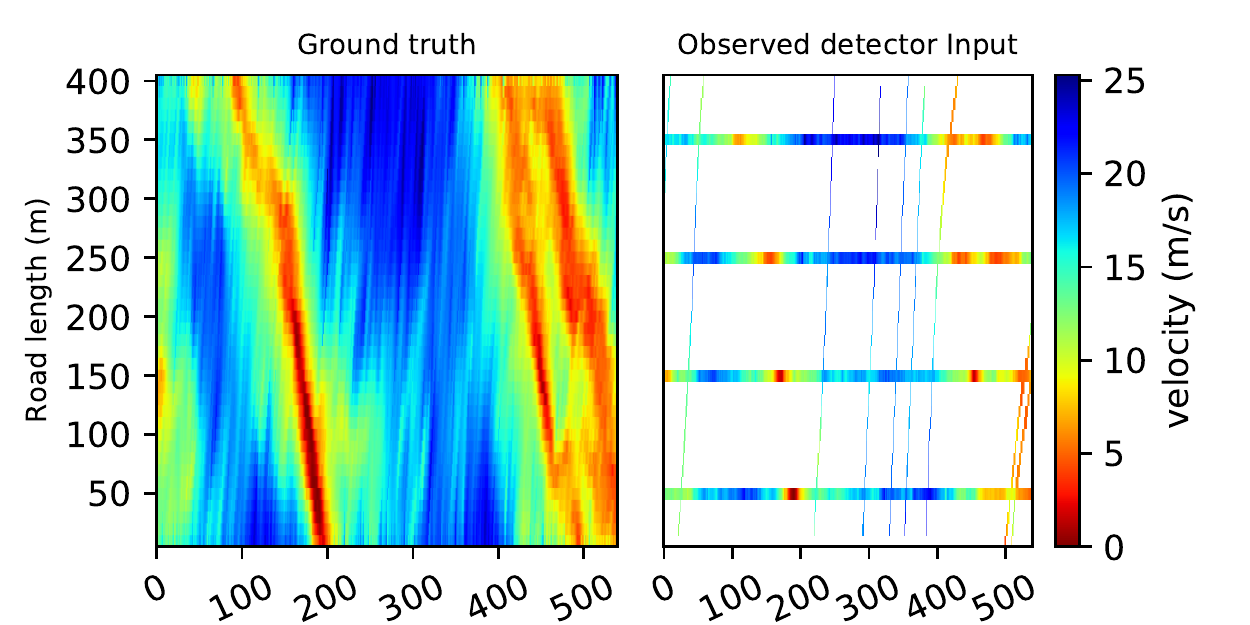}    
\caption{German highway data with heterogeneous inputs} 
\label{fig:germandata}
\end{center}
\end{figure}

We test the performance of our method using the NGSIM US 101 highway data and German highway data.
The NGSIM US 101 data consists of vehicle trajectories from a road section which is 670 meters in length and German highway data consists of vehicle trajectories from a 400 meter road section. They are converted to a ground-truth macroscopic speed field $\mathbf{V(x,t)}$. Thus in this work, we deal with estimating the speed field.
Heatmaps of the speed fields are shown in Fig.~\ref{fig:usdata} and Fig.~\ref{fig:germandata} for NGSIM US 101 and German highway data, respectively, both over an interval of length 600 seconds. The US101 data come from 4 evenly spaced stationary traffic sensors (inductance loop detectors). Input of German highway data come from a combination of $5\%$ floating car trajectories and 2 stationary detectors. Both cases involve complex build-up and dissipation of congestion with mixtures of free flow, congested traffic, and shock-waves traveling in the opposite direction of traffic (the red bands in the figures).

We conduct two kinds of experiments in this study. In the first experiment, we compare the estimation error of the proposed ASNN algorithm and that of the conventional ASM. We use the typical values from Table \ref{table_parameter} as optimal settings for ASM and as initialization of ASNN. The ASNN is trained is using both convolution cost function and the physics-informed cost functions. 
 In the second experiment, we evaluate the benefits of the proposed MASNN estimation algorithm
 when the data is from heterogeneous sources where a good initialization of the ASNN parameters wasn't known. 

We optimize $\lambda$ using a grid search on candidate values $[10,5,1,0.5,0.1,0.05,0.01]$ and set $\lambda$ to 0.01 in both experiments. The kernel $\omega$ in the convolution cost function is chosen as
\begin{equation}
\label{eqn:kernel}
\begin{aligned}
\omega = \begin{bmatrix}
-1 & 0 & 0\\
-1 & 3 & 0 \\
-1 & 0 & 0
\end{bmatrix}
\end{aligned}
\end{equation}
The heuristic behind this kernel is that current traffic states are influenced by past states.

The traffic density $\rho$  introduced in the physics-informed cost function is calculated by inverting the Newell-Franklin relation \cite{newell1961nonlinear}, $v = V(\rho) = \frac{1}{\rho}Q(\rho)$, to transform speed value in each cell to a density as follows:
\begin{equation}
    \rho(x,t) = V^{-1}\big( \rho(x,t) \big) =  \frac{\rho_{\rm jam}}{1-\frac{v_{\rm f}}{|c_{\rm cong}|}\ln{(1-\frac{v(x,t)}{v_{\rm f}})}},
\end{equation}
where $\rho_{\rm jam}$ is the jam density (roughly 120 veh/km/lane), $v_{\rm f}$ is the desired (free-flow) speed, which is comparable to the speed limit of the road. We set $v_{\rm f} = 100$ km/h,  which is greater than maximum value of speed data observed, and $c_{\rm cong}$ is the congested wave speed. With this set of parameters, $\rho_{\rm cr}$ is 30 veh/km/lane. 

We investigate the algorithm’s accuracy numerically and by reconstructing the resulting heatmap visually. The estimation quality of the speed field $\widehat{\mathbf{V}}$ is measured by the relative error in all experiments:
\begin{equation}
\begin{aligned}
    m_{\mathrm{r}} = \frac{\|\widehat{\mathbf{Z}} - \mathbf{Z}\|_{\mathrm{F}}}{\|\mathbf{Z}\|_{\mathrm{F}}}
\end{aligned}
\end{equation}


\section{RESULT ANALYSIS AND DISCUSSION}\label{sec:discuss}

\textit{Expt 1: Comparison of ASM and ASNN:}
\newline
The estimation errors $m_{\mathrm{r}}$ for conventional ASM and ASNN with both convolution and physics losses are summarized in Table \ref{table_result}, with best performance highlighted in bold. 
\begin{table}[h]
\caption{Estimation errors results}
\label{table_result}
\begin{center}
\begin{tabular}{|c|c||c|c|}
\hline
US101 & $m_{\mathrm{r}}$  & Germany & $m_{\mathrm{r}}$ \\
\hline
ASM & 0.12417 & ASM & 0.08179 \\
ASNN (Conv) & 0.11868 & ASNN (Conv) & \textbf{0.08123} \\
ASNN (Phys) & \textbf{0.11866} & ASNN (Phys) & \textbf{0.08123} \\
\hline
\end{tabular}
\end{center}
\end{table}

\begin{figure}[h!]
\begin{center}
\includegraphics[width=.5\textwidth]{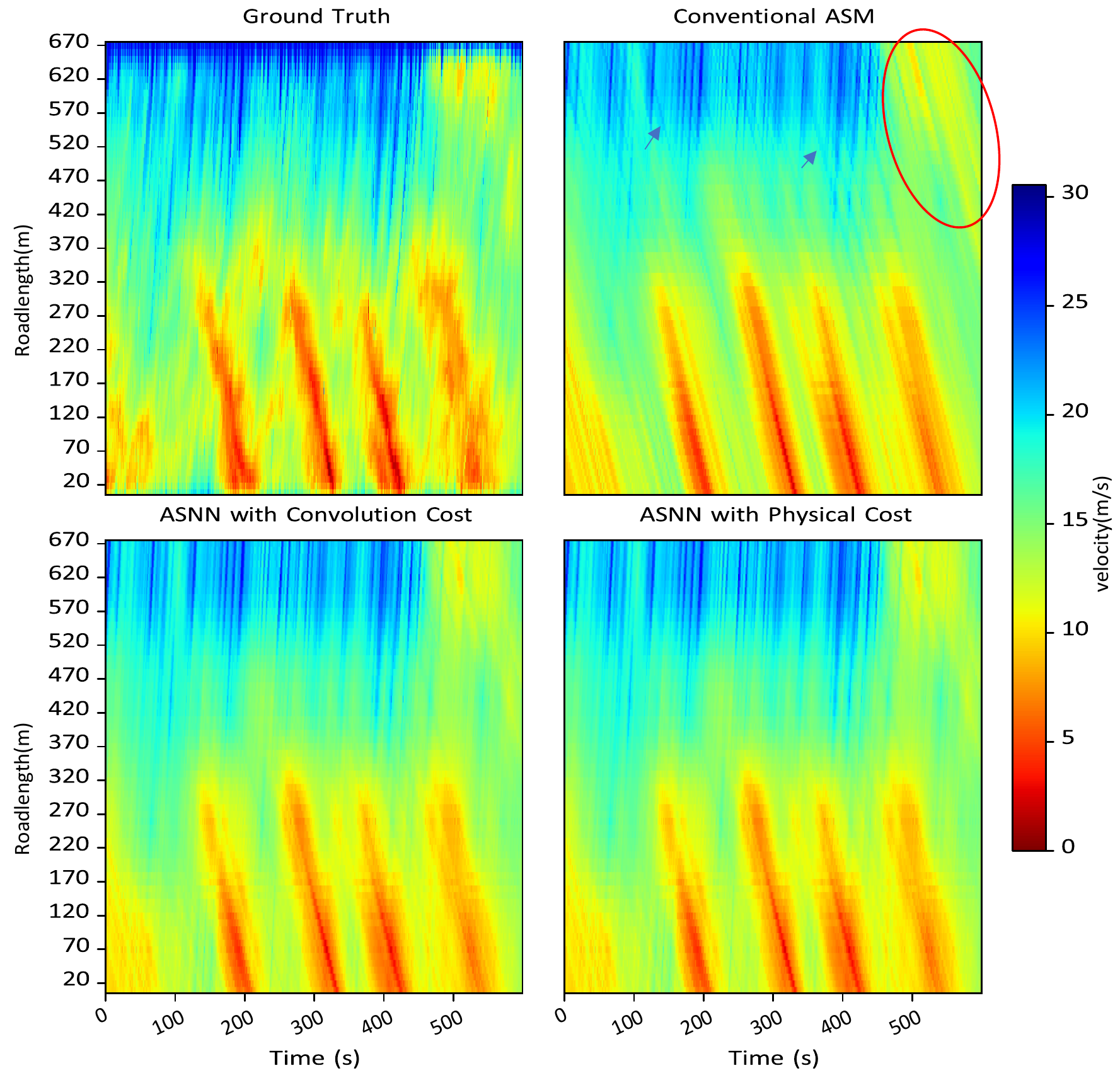}    
\caption{Estimation results for US101 data with stationary detector inputs: In the pane showing the conventional ASM case, non-physical patterns are highlighted with dark arrows and a red circle.} 
\label{fig:US101_result}
\end{center}
\end{figure}

We observe that the ASNN estimations have better performance over ASM in both cases.  While the error estimates of ASM and ASNN don't differ significantly, we note that the conventional ASM predictions are non-physical for US 101 data when the typical parameter values in Table \ref{table_parameter} are used. Specifically, we observe a cross-hatch in the ASM estimates when in free-flow, which violates causality in traffic (free-flowing traffic propagates with the direction of travel of the vehicles). The ASNN corrected the non-physical predictions produced by ASM (see Fig.\ref{fig:US101_result}). 

\begin{figure}[h!]
\begin{center}
\includegraphics[width=.5\textwidth]{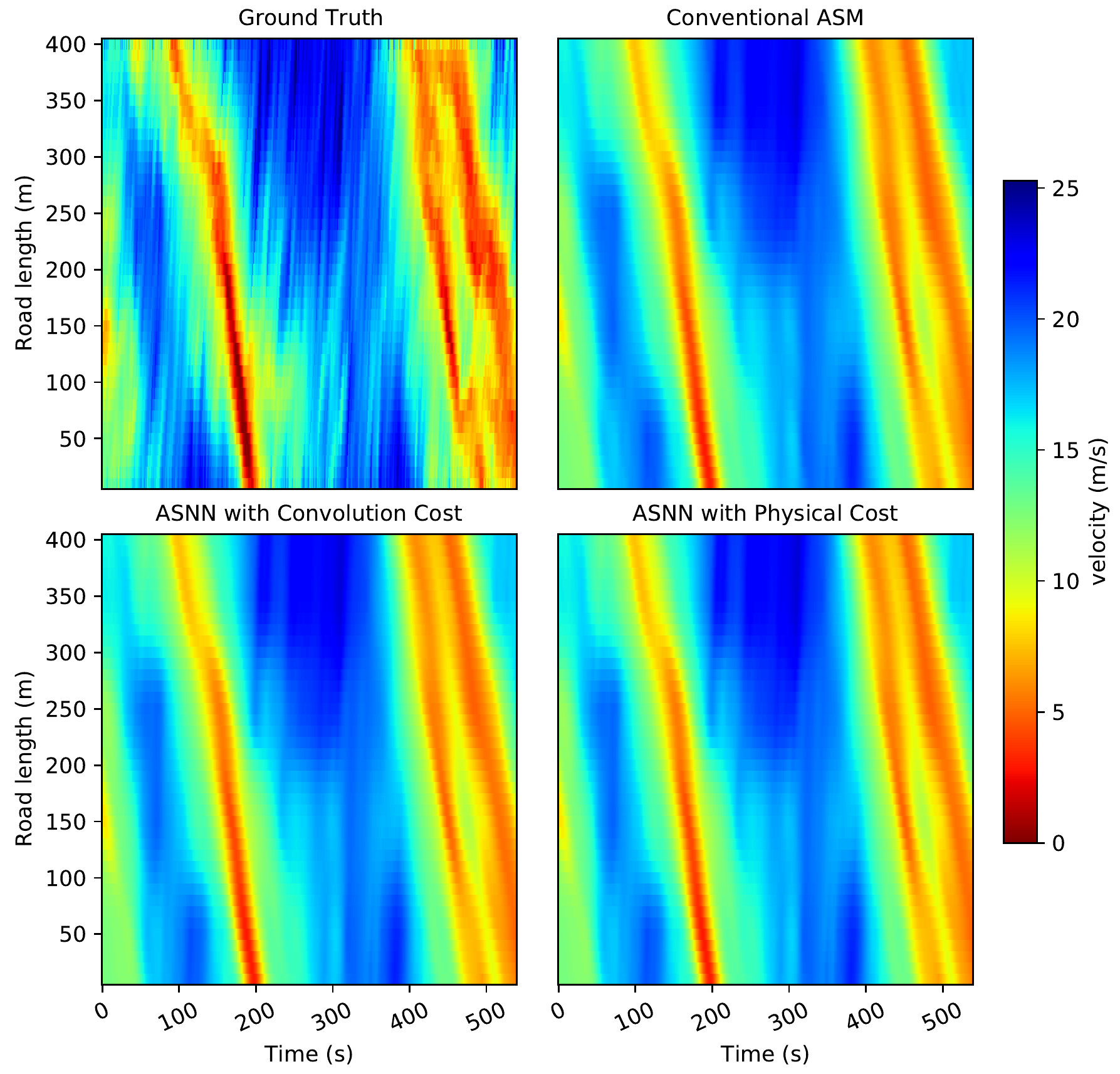}    
\caption{Estimation results for German highway data with heterogeneous inputs} 
\label{fig:germany_result}
\end{center}
\end{figure}

We note that as the data available for interpolation increases, the ASM gives more accurate predictions and its accuracy gets closer to that of ASNN. For instance, we used trajectory data in addition to detector data to estimate the traffic state for the German highway and noted that the ASM results are closer to ASNN results. See Fig. \ref{fig:germany_result}.

ASNNs trained using convolution cost and the physics cost perform equally well in this case.
 This indicates that temporal causality considered in the convolution cost function is effective in reconstructing traffic patterns. It also suggests that it is reasonable to adopt traffic flow models as regularization terms if we have prior information on the traffic patterns to be reconstructed. 
We noted in our experiments that the ASNN converges well if the initialization parameter values are chosen to be within the range between 0 and their typical values in Table \ref{table_parameter}.

\textit{Expt 2: MASNN with multiple a priori estimates}

Here, we discuss the results we obtained using  5 sets of possible initializations for the German highway data with 
heterogeneous inputs. We use 5 different initialization values for $\{\tau: 2.5,5,7.5,10,12.5 \}$. The congested and free-flow wave speeds are fixed at -15 km/hr and 80 km/hr, respectively for each initialization set. $V_{\rm thr}$ and $\Delta v$ are initialized to 1 km/hr in all the sets. The $\sigma$ is still considered to be $\Delta x/2$ i.e., half the distance between the detectors. 
The five different initializations for $\tau$ were chosen because of uneven time gaps in the floating car data. We note that adding more initializations doesn't change the results significantly.

\begin{figure}[h!]
\begin{center}
\includegraphics[width=.5\textwidth]{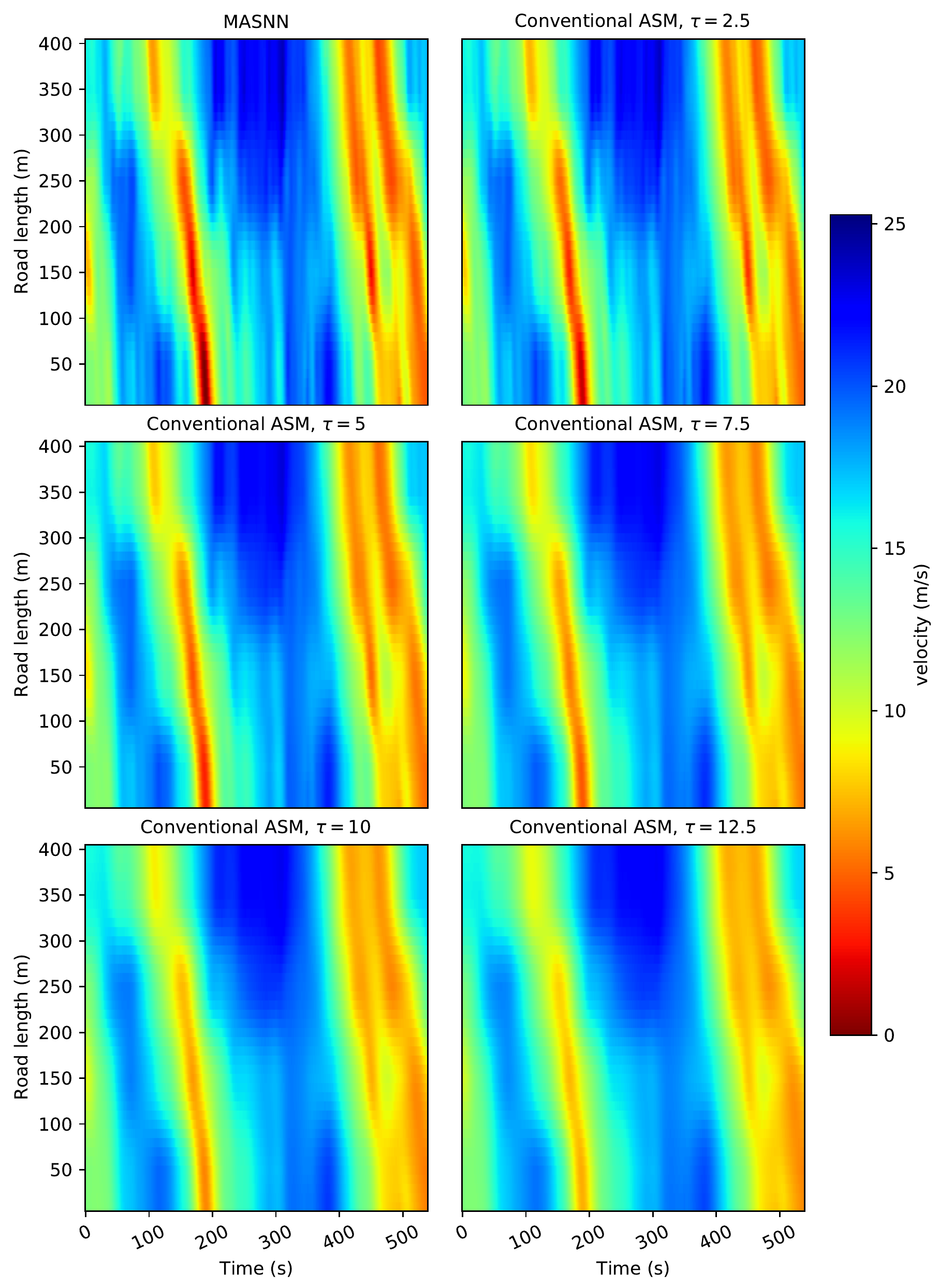}    
\caption{MASNN estimation results and its optimal temporal smoothing widths applied in ASM} 
\label{fig:MASNN}
\end{center}
\end{figure}

The estimation accuracy $m_{\mathrm{r}}$ of MASNN is 0.08021.  We note from Table \ref{table_result} and also visually observe from the Fig.\ref{fig:germany_result} and Fig.\ref{fig:MASNN}, MASNN's ability to combine multiple a priori estimates to improve its performance over that of ASNN.

The trained weights in the convex combination layer are recorded in Table \ref{table_weight}. 
\begin{table}[h]
\caption{MASNN trained weights in convex combination layer}
\label{table_weight}
\begin{center}
\begin{tabular}{|c|c|c|}
\hline
 Germany: $\sigma,\tau$ & weight & ASM $m_{\mathrm{r}}$\\
\hline
\textbf{50,2.5} & \textbf{1} & \textbf{0.08240}\\
 \textbf{50,5} & \textbf{1} & \textbf{0.08674}\\
 50,7.5 & 0 & 0.09367\\
 50,10 & 0 & 0.10154\\
 50,12.5 & 0 & 0.10963\\
\hline
\end{tabular}
\end{center}
\end{table}
Each weight setting is then tested with conventional ASM estimation and its corresponding estimation error is recorded. The initialization set with highest trained weight is highlighted in bold. A comparison of the corresponding ASM error reveals that MASNN preferred weights that produce better estimation results for ASM, which indicates that MASNN is also capable of choosing suitable parameters for conventional ASM. Fig.~\ref{fig:MASNN} displays a heatmap of the speed field produced by MASNN and ASM with different parameter settings. 
We observe that MASNN produced better shockwave patterns in comparison to the ground truth visualization than those estimated by ASM and ASNN in Fig.~\ref{fig:germany_result}. Even with the parameter set(s) determined by MASNN, the ASM has smoothened out the sharp changes more than the MASNN. The smoothening effect in ASM increases with increase in $\tau$.

\section{SUMMARY AND CONCLUSION}\label{sec:conclusion}
In this study, we develop a neural network (ASNN) based on adaptive smoothing method (ASM). The ASNN 
tunes the parameters of the ASM by sparse data from detectors and reconstructs the traffic state all along the road. We tested the ASNN with two cost functions. One 
minimizes data loss with a regularizer that penalizes sharp differences in speed in neighboring cells.
The other cost function minimizes the data loss with a  physics-informed regularizer based on the principle of conservation of traffic density. 
We also proposed a way to enhance the robustness of the ASNN in learning by using an ensemble averaging technique which we refer as MASNN. We applied the ASNN and the MASNN on two real-world data
sets; one is the NGSIM US 101 highway data set which is from stationary detectors, the other uses German highway data from heterogeneous sources including stationary detectors and floating vehicles.
Our experiments reveal that ASNN initialized with typical parameter values used for ASM outperforms the ASM in estimating the traffic state in all the cases. In the case of heterogeneous detectors where  a good initialization set for parameters is not known, the MASNN has shown excellent performance in estimating traffic. We also note that the initialization parameters with the highest weights as predicted by MASNN, can serve as good parameter values for conventional ASM. Overall, we find that the MASNN and the ASNN are promising tools for traffic state estimation.


We find scope for future improvements:
The ASNN architecture suffers from the vanishing gradient effect which affects the the parameters in the initial layers. Moreover, the ASM being a smoothing method, it smears out sharp variations in the patterns. A feed-forward network, in-lieu of the smoothing layer might improve the performance in both the above-mentioned aspects. The use of Frobenius-norm based costs focused on large numbers, which results in inaccurate estimates when the speeds are low. One can circumvent this by using density as the variable to be estimated or adding alternate penalty terms to emphasize low-speed traffic.

\section*{Acknowledgment}
This work was supported by the NYUAD Center for Interacting Urban Networks (CITIES), funded by Tamkeen under the NYUAD Research Institute Award CG001.  The opinions expressed in this article are those of the authors alone do not represent the opinions of CITIES.

	
	
{ \small
\bibliographystyle{plainnat}
\bibliography{References}

\begin{thebibliography}{23}
\providecommand{\natexlab}[1]{#1}
\providecommand{\url}[1]{\texttt{#1}}
\expandafter\ifx\csname urlstyle\endcsname\relax
  \providecommand{\doi}[1]{doi: #1}\else
  \providecommand{\doi}{doi: \begingroup \urlstyle{rm}\Url}\fi

\bibitem[Aw and Rascle(2000)]{aw2000resurrection}
AATM Aw and Michel Rascle.
\newblock Resurrection of" second order" models of traffic flow.
\newblock \emph{SIAM journal on applied mathematics}, 60\penalty0 (3):\penalty0
  916--938, 2000.

\bibitem[Benkraouda et~al.(2020)Benkraouda, Thodi, Yeo, Menendez, and
  Jabari]{benkraouda2020traffic}
Ouafa Benkraouda, Bilal~Thonnam Thodi, Hwasoo Yeo, Monica Menendez, and
  Saif~Eddin Jabari.
\newblock Traffic data imputation using deep convolutional neural networks.
\newblock \emph{IEEE Access}, 8:\penalty0 104740--104752, 2020.

\bibitem[Chen et~al.(2018)Chen, Zhang, Li, and Li]{chen2018adaptive}
Xiqun Chen, Shuaichao Zhang, Li~Li, and Liang Li.
\newblock Adaptive rolling smoothing with heterogeneous data for traffic state
  estimation and prediction.
\newblock \emph{IEEE transactions on intelligent transportation systems},
  20\penalty0 (4):\penalty0 1247--1258, 2018.

\bibitem[Huang and Agarwal(2020)]{huang2020physics}
Jiheng Huang and Shaurya Agarwal.
\newblock Physics informed deep learning for traffic state estimation.
\newblock In \emph{2020 IEEE 23rd International Conference on Intelligent
  Transportation Systems (ITSC)}, pages 1--6. IEEE, 2020.

\bibitem[Jabari et~al.(2018{\natexlab{a}})Jabari, Dilip, Lin, and
  Thodi]{jabari2018learning}
Saif~Eddin Jabari, Deepthi~Mary Dilip, DianChao Lin, and Bilal~Thonnam Thodi.
\newblock Learning traffic flow dynamics using random fields.
\newblock \emph{arXiv preprint arXiv:1806.08764}, 2018{\natexlab{a}}.

\bibitem[Jabari et~al.(2018{\natexlab{b}})Jabari, Zheng, Liu, and
  Filipovska]{jabari2018stochastic}
Saif~Eddin Jabari, Fangfang Zheng, H~Liu, and Monika Filipovska.
\newblock Stochastic lagrangian modeling of traffic dynamics.
\newblock In \emph{Proc. 97th Annu. Meeting Transp. Res. Board}, pages 1--14,
  2018{\natexlab{b}}.

\bibitem[Jabari et~al.(2020)Jabari, Freris, and Dilip]{jabari2020sparse}
Saif~Eddin Jabari, Nikolaos~M Freris, and Deepthi~Mary Dilip.
\newblock Sparse travel time estimation from streaming data.
\newblock \emph{Transportation Science}, 54\penalty0 (1):\penalty0 1--20, 2020.

\bibitem[Jia et~al.(2016)Jia, Wu, and Du]{jia2016traffic}
Yuhan Jia, Jianping Wu, and Yiman Du.
\newblock Traffic speed prediction using deep learning method.
\newblock In \emph{2016 IEEE 19th international conference on intelligent
  transportation systems (ITSC)}, pages 1217--1222. IEEE, 2016.

\bibitem[LeVeque and Leveque(1992)]{leveque1992numerical}
Randall~J LeVeque and Randall~J Leveque.
\newblock \emph{Numerical methods for conservation laws}, volume 214.
\newblock Springer, 1992.

\bibitem[Li et~al.(2022)Li, Yang, and Jabari]{li2022nonlinear}
Wenqing Li, Chuhan Yang, and Saif~Eddin Jabari.
\newblock Nonlinear traffic prediction as a matrix completion problem with
  ensemble learning.
\newblock \emph{Transportation science}, 56\penalty0 (1):\penalty0 52--78,
  2022.

\bibitem[Lighthill and Whitham(1955)]{lighthill1955kinematic}
Michael~James Lighthill and Gerald~Beresford Whitham.
\newblock On kinematic waves ii. a theory of traffic flow on long crowded
  roads.
\newblock \emph{Proceedings of the Royal Society of London. Series A.
  Mathematical and Physical Sciences}, 229\penalty0 (1178):\penalty0 317--345,
  1955.

\bibitem[Newell(1961)]{newell1961nonlinear}
Gordon~Frank Newell.
\newblock Nonlinear effects in the dynamics of car following.
\newblock \emph{Operations research}, 9\penalty0 (2):\penalty0 209--229, 1961.

\bibitem[Richards(1956)]{richards1956shock}
Paul~I Richards.
\newblock Shock waves on the highway.
\newblock \emph{Operations research}, 4\penalty0 (1):\penalty0 42--51, 1956.

\bibitem[Schreiter et~al.(2010)Schreiter, van Lint, Treiber, and
  Hoogendoorn]{schreiter2010two}
Thomas Schreiter, Hans van Lint, Martin Treiber, and Serge Hoogendoorn.
\newblock Two fast implementations of the adaptive smoothing method used in
  highway traffic state estimation.
\newblock In \emph{13th International IEEE Conference on Intelligent
  Transportation Systems}, pages 1202--1208. IEEE, 2010.

\bibitem[Seo et~al.(2017)Seo, Bayen, Kusakabe, and Asakura]{seo2017traffic}
Toru Seo, Alexandre~M Bayen, Takahiko Kusakabe, and Yasuo Asakura.
\newblock Traffic state estimation on highway: A comprehensive survey.
\newblock \emph{Annual reviews in control}, 43:\penalty0 128--151, 2017.

\bibitem[Shi et~al.(2021)Shi, Mo, Huang, Di, and Du]{shi2021physics}
Rongye Shi, Zhaobin Mo, Kuang Huang, Xuan Di, and Qiang Du.
\newblock A physics-informed deep learning paradigm for traffic state and
  fundamental diagram estimation.
\newblock \emph{IEEE Transactions on Intelligent Transportation Systems}, 2021.

\bibitem[Thodi et~al.(2021)Thodi, Khan, Jabari, and
  Men{\'e}ndez]{thodi2021learning}
Bilal~Thonnam Thodi, Zaid~Saeed Khan, Saif~Eddin Jabari, and M{\'o}nica
  Men{\'e}ndez.
\newblock Learning traffic speed dynamics from visualizations.
\newblock In \emph{2021 IEEE International Intelligent Transportation Systems
  Conference (ITSC)}, pages 1239--1244. IEEE, 2021.

\bibitem[Thodi et~al.(2022)Thodi, Khan, Jabari, and
  Menéndez]{thodi2022incorporating}
Bilal~Thonnam Thodi, Zaid~Saeed Khan, Saif~Eddin Jabari, and Monica Menéndez.
\newblock Incorporating kinematic wave theory into a deep learning method for
  high-resolution traffic speed estimation.
\newblock \emph{IEEE Transactions on Intelligent Transportation Systems},
  23\penalty0 (10):\penalty0 17849--17862, 2022.

\bibitem[Treiber and Helbing(2002)]{treiber2002reconstructing}
Martin Treiber and Dirk Helbing.
\newblock Reconstructing the spatio-temporal traffic dynamics from stationary
  detector data.
\newblock \emph{Cooper@ tive Tr@ nsport@ tion Dyn@ mics}, 1\penalty0
  (3):\penalty0 3--1, 2002.

\bibitem[Treiber et~al.(2011)Treiber, Kesting, and
  Wilson]{treiber2011reconstructing}
Martin Treiber, Arne Kesting, and R~Eddie Wilson.
\newblock Reconstructing the traffic state by fusion of heterogeneous data.
\newblock \emph{Computer-Aided Civil and Infrastructure Engineering},
  26\penalty0 (6):\penalty0 408--419, 2011.

\bibitem[van Wageningen-Kessels(2019)]{van2019traffic}
Femke van Wageningen-Kessels.
\newblock Traffic flow modelling: Introduction to traffic flow theory through a
  genealogy of models.
\newblock 2019.

\bibitem[Xiao et~al.(2018)Xiao, Wei, and Liu]{xiao2018speed}
Jianli Xiao, Chao Wei, and Yuncai Liu.
\newblock Speed estimation of traffic flow using multiple kernel support vector
  regression.
\newblock \emph{Physica A: Statistical Mechanics and its Applications},
  509:\penalty0 989--997, 2018.

\bibitem[Yang et~al.(2022)Yang, Thodi, and Jabari]{yang2022generalized}
Chuhan Yang, Bilal~Thonnam Thodi, and Saif~Eddin Jabari.
\newblock Generalized adaptive smoothing using matrix completion for traffic
  state estimation.
\newblock \emph{arXiv preprint arXiv:2206.01461}, 2022.

\end{thebibliography}
}
	
	
	
	
	

\end{document}